\documentclass[12pt]{article}
\def\fslash#1{\ooalign{\hfil/\hfil\crcr$#1$} }
\usepackage{a4}
\usepackage{amsmath}
\usepackage{amssymb}
\usepackage[dvips]{graphicx}
\begin{document}
%%%%%%%%%%%%%%%%%%%%%%%%%%%%%%%%%%%%%%%%%%%%%%%%

\begin{flushright}
{UT-Komaba 01-07}
\end{flushright}
\vskip 0.5 truecm
%%%%%%%%%%%%%%%%%%%%%%%%%%%%%%%%%%%%%%%%%%%%%%%%

%%%%%%%%%%%%%%%%%%%%%%%%%%%%%%%%%%%%%%%%%%%%%%%%

\begin{center} 
{\Large{\bf Vortex fermion on the lattice \footnote{This is the revised version 
where we corrected the errors related to $k\neq0$ zero modes, which were pointed 
out by Neuberger in ref.\cite{Neuberger:2003yg}.}}}\\
\vskip 1.5cm

{\Large  Keiichi Nagao\footnote{Correspondence address: Faculty of Education, 
Ibaraki University, Mito 310-8512, Japan \\
E-mail address: nagao@mx.ibaraki.ac.jp }}  
\vskip 1.0cm
 
{\it Institute of Physics, University of Tokyo, 
%3-8-1 
Komaba, 
%Meguro-ku, 
Tokyo 153-8902, Japan  }\\

%{\it and }\\

{\it 
Institute of Particle and Nuclear Studies, 
%High Energy Accelerator Research Organization (
KEK, 
%1-1 Oho, 
Tsukuba 
%Ibaraki 
305-0801, Japan}\\

%{\it and }\\

{\it 
%Theoretical Physics Laboratory,
Faculty of Education, Ibaraki University, 
%2-1-1 Bunkyo, 
Mito 
%Ibaraki 
310-8512, Japan  }\\

%{\it and }\\

{\it Niels Bohr Institute, 
Blegdamsvej 17, DK2100 Copenhagen, Denmark}\\

\end{center}

\vskip 1.5cm
\begin{center} 
\begin{bf}
Abstract
\end{bf}
\end{center}

The domain wall fermion formalism in lattice gauge theory  
is much investigated recently.
This is set up by reducing 
$4+1$ dimensional theory to low energy effective $4$ 
dimensional one.
In order to look around other possibilities of realizing 
chiral fermion on the lattice,
we construct vortex fermion by reducing $4+2$ 
dimensional theory to 
low energy effective $4$ dimensional one on the lattice.
In extra $2$ dimensions we propose a {\it new} lattice 
regularization 
which has a discrete {\it rotational} invariance but not a 
translational one.
In order to eliminate doubling species in the naive construction 
we introduce 
the extended Wilson term 
which is appropriate to our model.
We propose two models for convenience and 
show that a normalizable zero mode solution 
appears at the core of the vortex.

\newpage
\setcounter{footnote}{0}

\section{Introduction}

In lattice gauge theory the domain wall fermion 
formalism\cite{Kaplan}\cite{Shamir} is much 
investigated recently.
Though domain wall fermion itself includes unwanted heavy 
modes in the bulk, the subtraction of them was 
discussed\cite{Kaplan2}\cite{subtraction}.
Subtracting the heavy modes in the case of 
the vector-like model, Neuberger succeeded to obtain the 
so-called overlap Dirac operator\cite{Neuberger} 
which satsifies G-W relation\cite{GinspargWilson}.
This enables us to have L\"uscher's extended chiral 
symmetry on the lattice\cite{Luscher}\cite{Nieder} 
and to circumvent 
Nielsen-Ninomiya's theorem\cite{Nielsen}. 
The well-known index theorem in continuum theory\cite{Atiya} 
is also realized on the lattice\cite{Hasenfratzindex}\cite{Luscher}.
L\"uscher succeeded in constructing abelian chiral 
gauge theory on the lattice\cite{abeliangauge}.
Domain wall fermion thus seems to provide a promising new 
definition for constructing chiral gauge theories 
on the lattice.

A key basis for this development is the observation 
that, in the presence of the mass defect which is 
introduced as a scalar background in higher $4+x$ dimensions, 
a chiral fermion zero mode appears at 
the defect\cite{RubSha}\cite{Callan}. 
The topological defects are 
a kink (domain wall) for $x=1$, 
a vortex for $x=2$, 
a monopole for $x=3$, and an instanton for $x=4$, 
respectively.
On the basis above Kaplan put the domain wall model, 
corresponding to $x=1$, 
on the lattice by introducing Wilson term\cite{Wilsonterm} 
and opened a road to 
construct the $4$ dimensional chiral gauge theory 
on the lattice\cite{Kaplan}.

In this paper we shall put the vortex fermion model, 
corresponding to $x=2$, on the lattice.
This is the higher dimensional generalization of 
domain wall fermion on the lattice.
%

%%%%%%%%%%%%%%%%%%%%%%%%%%%%%%%%%%%%%%%%%%%%%%%%%%%%%
We have several motivations to construct vortex fermion 
on the lattice.
One among them is that it is interesting in itself to generalize 
domain wall fermion in $5$ dimensions 
to vortex fermion in $6$ dimensions.
The most important property of domain wall fermion is that 
overlap Dirac operator\cite{Neuberger}, which was derived by 
subtracting unwanted modes from vector-like 
domain wall fermion, 
satisfies G-W relation\cite{GinspargWilson}. 
It is, however, remained unknown why those two different ideas 
(extra-dimension and renormalization group methods) are linked, 
though it is considered how G-W relation is broken 
at finite $a_5, N_5$ in domain wall fermion\cite{kikunogu}.  

In order to investigate this and to obtain another formalism 
of chiral fermion on the lattice, it is interesting 
and worthwhile to derive a new Dirac operator which 
will be obtained by subtracting unwanted 
modes from vector-like vortex fermion 
and to check whether this new Dirac operator 
satisfies G-W relation or not.
If it satisifies G-W relation, it means that we have got 
a new solution of G-W relation, and that we have developed 
the new formalism of chiral fermion on the lattice.
Furthermore we may obtain 
some information of the reason why two different ideas 
mentioned above are linked.
The work of this paper, though we construct only chiral 
model but not vector-like one, 
may be a preliminary step to proceed along such direction.

%%%%%%%%%%%%%%%%%%%%%%%%%%%%%%%%
On the other hand, when we discuss gauge anomaly in 
$4$ dimensions, $6$ dimensional physics is a key point, 
since it is known that 
$4$ dimensional gauge anomaly is related with $5$ dimensional 
Chern-Simon term and $6$ dimensional Dirac index density 
by algebraic\cite{Zumino} and 
topological\cite{AlvarezGins} explanations.
Furthermore 
Callan and Harvey explained that the $4$ dimensional gauge 
anomaly caused by the zero mode localized at the defect 
is understood  as just a charge flow in $4+x$ 
dimensions where the charge is conserved\cite{Callan}. 
They explicitly developed this picture in the case 
of $x=1,2$, and 
gave a physical explanation by fermion zero modes 
to gauge anomaly.
It hence makes sense to study vortex fermion 
from the view point of gauge anomaly on the lattice.
We think that the investigation of $6$ dimensional model 
is {\it as important as} that of $5$ dimensional one 
for the purpose of understanding the {\it whole} structure of 
gauge anomaly and constructing chiral gauge theories 
on the lattice. This is the principal motivation for 
considering $6$ dimensional model.

L\"uscher discussed constructing non-abelian chiral 
gauge theory with G-W fermion in $4+2$ dimensions 
where $2$ dimensional parameters are continuous, 
which interpolate gauge fields in the link variable 
space\cite{weylfermion} which are bounded for the plaquette 
$\Vert 1- U(p) \Vert < \epsilon$\cite{locality}\cite{kikulocal}. 
Kikukawa discussed this with domain wall 
fermion in $5+1$ dimensions where $1$ dimension is 
continuous\cite{kikudomain}.
The non-abelian anomaly is also discussed 
in $4+2$ dimensions where $2$ dimensions are 
continuous\cite{barneu}\cite{suzuki}\cite{adams}.
We propose to extend such possibilities to our vortex 
fermion in $6$ dimensions where whole $6$ dimensions 
are discretized.\footnote{After the previous version of this paper was submitted, 
Neuberger studied the vortex fermion where extra 2 dimensions are continuous in 
ref.\cite{Neuberger:2003yg}.}
%%%%%%%%%%%%%%%%%%%%%%%%%%%%%%%%%%%

%%%%%%%%%%%%%%%%%%%%%%%%%%%%%%%%%%%%%%%%%
For this purpose, we first study vortex fermion in 
continuum theory in section 2 where we note that 
the {\it rotational} invariance 
in extra $2$ dimensions has an important role.
In section 3 we shall construct vortex fermion on the lattice. 
The usual square (cubic) lattice is not appropriate for 
this purpose since 
it has a discrete translational invariance 
but not a rotational one.
Therefore we define and propose a {\it new} 
lattice regularization 
which is obtained by discretizing 
polar coordinates in extra $2$ dimensions.
This {\it new} reguralization, ``{\it spider's web}'' 
lattice, which has a discrete {\it rotational} invariance 
but not a translational 
one, is one of characteristics of this paper.
Using this spider's web lattice, we shall 
construct vortex fermion on the lattice. 
The action, which is constructed so that they have 
the hermiticity in both $\rho$ and $\phi$ directions, 
includes a free parameter which should be fixed 
appropriately. This indicates some ambiguity of the 
definition at the string, namely the core of the vortex.
We impose some constraint on the action to avoid the ambiguity 
and another method of parameter fixing is discussed in 
appendix. The naive construction of vortex fermion 
suffers from the appearance of doubling species.
In order to eliminate them, 
we construct the extended Wilson term which is modified to 
include scalar fields to be appropriate to our model. 
Using this extended Wilson term we show that there appears a 
normalizable zero mode 
solution localized at the string.
For convenience, we propose two models.
One is a simpler model which makes easier for us to confirm 
the appearance of a normalizable zero mode solution, though 
this model does not have the hermiticity in the 
$\rho$ direction at a {\it finite} lattice spacing $a_\rho$. 
Another is an hermitian model which is the main result of 
this paper. We show that the desired 
solution appears also in this more satisfactory model.
Finally we discuss the summary of this paper and 
the future work in section 4.
%
%%%%%%%%%%%%%%%%%%%%%%%%%%%%%%%%%%%%%%%%%%%%%%%%

\section{Vortex fermion in continuum theory}

We start from a brief review of the vortex fermion 
system in the continuum theory\cite{Callan}.

The model in $2n+2$ dimensional Euclidean space is given by the 
following Lagrangian,
\begin{eqnarray}
{\cal L}&=&\bar\psi \sum_{\mu=1}^{2n+2}
\Gamma_\mu \partial_\mu \psi + 
\bar\psi (\Phi_1 + i\bar\Gamma \Phi_2)\psi ,\\
\Phi &=& f(\rho)e^{i\phi}=\Phi_1 + i\Phi_2, \\
\langle\Phi\rangle &=& \nu.
\end{eqnarray}
where $\rho$ and $\phi$ are polar coordinates defined by 
\begin{equation}
x_{2n+1}=\rho \cos\phi,\quad x_{2n+2}=\rho \sin\phi.
\end{equation}
$\Phi$ is a fixed classical complex scalar 
field\footnote{The phase of $\Phi$ is associated 
with the axion field 
$\theta(x)$ and we consider the situation that the 
string runs along the $(x_1,\cdots, x_{2n})$-axis, 
perpendicular to $(x_{2n+1},x_{2n+2})$-plane so that we 
can take the axion field $\theta(x)=\phi$ in 
cylindrical coordinates.}
and 
$f(\rho)$ is an increasing function of $\rho$ which satisfies 
$f(0)=0$ 
and which approaches $\nu$ at infinity. 
$\bar\Gamma$ is defined as follows
\begin{eqnarray} 
\bar\Gamma &=& \Gamma_{int}\Gamma_{ext}, \\
\Gamma_{int}&=&(-i)^{n}\Gamma_1\cdots\Gamma_{2n} , \\
\Gamma_{ext}&=&(-i)\Gamma_{2n+1}\Gamma_{2n+2} , 
\end{eqnarray}
where $\Gamma_\mu (\mu=1,2,\cdots , 2n+2)$ are 
$2n+2$ dimensional hermitian 
gamma matrices\footnote{Explicit notation 
of gamma matrices is given in appendix section.} and 
$\Gamma_{int}$ is the chiral operator on the string.

Then the Dirac equation is given by 
\begin{eqnarray}
&&\sum_{\mu=1}^{2n+2}\Gamma_\mu \partial_\mu \psi \nonumber \\
&=&
\left[\sum_{i=1}^{2n}\Gamma_i \partial_i +
\Gamma_{2n+1}(\cos\phi +i\Gamma_{ext}\sin\phi)\partial_\rho -
\Gamma_{2n+1}(\sin\phi -i\Gamma_{ext}\cos\phi)
\frac{1}{\rho}\partial_\phi \right]\psi \nonumber\\
&=&
-(\Phi_1 + i\bar\Gamma \Phi_2)\psi, 
\end{eqnarray}
which is equivalent to 
\begin{eqnarray}
\sum_{i=1}^{2n}\gamma_i \partial_i \psi_{\beta}+
\Gamma_{2n+1}(\cos\phi +i\Gamma_{ext}\sin\phi)
\partial_\rho \psi_{\beta} 
&-&\Gamma_{2n+1}(\sin\phi -i\Gamma_{ext}\cos\phi)
\frac{1}{\rho} \partial_\phi \psi_{\beta} \nonumber \\
 &=&-f(\rho)e^{i\phi}\psi_{\alpha} , \\
\sum_{i=1}^{2n}\gamma_i \partial_i \psi_{\alpha}+
\Gamma_{2n+1}(\cos\phi +i\Gamma_{ext}\sin\phi)
\partial_\rho \psi_{\alpha} 
&-&\Gamma_{2n+1}(\sin\phi -i\Gamma_{ext}\cos\phi)
\frac{1}{\rho} \partial_\phi \psi_{\alpha} \nonumber \\
&=&-f(\rho)e^{-i\phi}\psi_{\beta} 
\end{eqnarray}
where $\psi_{\alpha}$ and $\psi_{\beta}$ are the eigenfunctions 
of $\bar\Gamma$,
\begin{equation} 
\bar\Gamma\psi_{\alpha}=\psi_{\alpha}, \quad 
\bar\Gamma\psi_{\beta}=-\psi_{\beta}.
\end{equation}

%%%%%%%%%%%%%%%%%%%%%%%%%%%%%%%%%%%%%%%%%%%%%%%%%%%%%%%%%%%%%%%%%%%%%%%%

We now solve the zero mode solution which satisfies 
\begin{equation}
\sum_{\mu=2n+1}^{2n+2}\Gamma_\mu \partial_\mu \psi 
+(\Phi_1 + i\bar\Gamma \Phi_2)\psi =0.\label{zeroc}
\end{equation}
We assume the following form for $\psi$
\begin{eqnarray}
\psi^{\pm}_\beta &=&e^{ip\cdot x}\varphi_{\pm}(\rho, \mbox{\boldmath $p$}, k) e^{ik\phi}
u_{\pm},\label{kbeta}\\
\psi^{\pm}_\alpha &=& \Gamma_{2n+1}
\psi^{\pm}_{\beta}\label{kalpha},
\end{eqnarray}
where $u_{\pm}$ are constant spinors which have a definite 
chirality respectively as $\Gamma_{int}u_{\pm}=\pm u_{\pm}$, 
and $k$ represents the angular-momentum. 
Substituting eqs.(\ref{kbeta})(\ref{kalpha}) into 
eq.(\ref{zeroc}), we obtain 
\begin{eqnarray}
e^{\mp i\phi}\left(e^{ik\phi}\partial_\rho 
\varphi_\pm (\rho) \mp i\frac{1}{\rho} \varphi_\pm(\rho) 
\partial_\phi e^{ik\phi}\right)
u_\pm &=&-f(\rho)e^{i\phi}\varphi_\pm(\rho)e^{ik\phi} u_\pm,\\
e^{\pm i\phi}\left(e^{ik\phi}\partial_\rho 
\varphi_\pm (\rho) \pm i\frac{1}{\rho} \varphi_\pm(\rho) 
\partial_\phi e^{ik\phi}\right)
u_\pm &=&-f(\rho)e^{-i\phi}\varphi_\pm(\rho)e^{ik\phi} u_\pm, \label{missed_eq}
\end{eqnarray}
where we have abbreviated $\varphi_{\pm}(\rho , \mbox{\boldmath $p$}, k)$ as 
$\varphi_{\pm}(\rho)$. 
These relations reduce to 
\begin{eqnarray}
\partial_\rho \varphi_-(\rho)-\frac{k}{\rho}\varphi_-(\rho)&=&
-f(\rho)\varphi_-(\rho),\\
\partial_\rho \varphi_-(\rho)+\frac{k}{\rho}\varphi_-(\rho)&=&
-f(\rho)\varphi_-(\rho). \label{missed_eq2}
\end{eqnarray}
The solution exists for $k=0$, and it is given by 
\footnote{In the previous version of this paper we missed eq.(\ref{missed_eq}), 
namely eq.(\ref{missed_eq2}), 
and reached the wrong conclusion that there are $k\neq 0$ zero modes besides 
a $k=0$ zero mode. 
Similar errors occurred also when 
we solved zero mode equations on the lattice 
%in the naive fermion model, the simple 
%vortex model and the hermitian vortex model 
in section 3, namely, in the naive fermion model, the simple vortex model and 
the hermitian vortex model. 
These errors were pointed out by Neuberger in 
ref.\cite{Neuberger:2003yg}. 
In this revised version all of the relevant errors are corrected 
and we see that there are no $k\neq0$ zero modes in each lattice model, which 
agrees with the result of ref.\cite{Neuberger:2003yg}.}
\begin{equation}
\varphi_-(\rho)=
\exp{\left(-\int_0^\rho f(\sigma)d\sigma \right)}. 
\end{equation}
This means that we have obtained a normalizable zero mode 
solution which has a definite chirality,  
\begin{eqnarray}
\psi^{-}_\beta &=&e^{ip\cdot x} 
\exp{\left(-\int_0^\rho f(\sigma)d\sigma \right)} 
u_{-}, \label{kneq0}\\
\psi^{-}_\alpha &=& \Gamma_{2n+1}\psi^{-}_{\beta}.
\end{eqnarray}
We note that the zero mode solution is localized along 
the string.

We briefly study the charge flow into the string in $2+2$ dimensions.
The expectation value of the current[26] 
%\cite{Goldstone} 
is given by 
\begin{equation}
\langle J_{\mu}\rangle=
-i\frac{e}{16\pi^2}\epsilon_{\mu\nu\lambda\rho}
\frac{(\Phi^* \partial^\nu\Phi-\Phi\partial^\nu\Phi^*)}
{\vert\Phi\vert^2}
F^{\lambda\rho}.
\end{equation}
We write $\Phi(x)$ as 
\begin{equation}
\Phi(x)=\nu e^{i\theta(x)}
\end{equation}
off the string using the axion field $\theta(x)$ 
and consider the situation that 
the string runs along the $x_1,x_2$-axis, perpendicular 
to the $x_3,x_4$-plane.
Using the following equation
\begin{equation}
[\partial_{x_3},\partial_{x_4}]\theta=
2\pi\delta(x_3)\delta(x_4)\label{theta}
\end{equation}
which represents the topology of the scalar field $\Phi$, 
we obtain 
\begin{equation}
\partial^\mu \langle J_\mu\rangle=
\frac{e}{4\pi}\epsilon^{ab} 
F_{ab}\delta(x_3)\delta(x_4), \quad a,b=1,2.
\end{equation}
We see that the anomaly arises at 
the string $(x_1,x_2,x_3,x_4)=(x_1,x_2,0,0)$, 
namely from the zero mode solution. 

%%%%%%%%%%%%%%%%%%%%%%%%%%%%%%%%%%%%%%%%%%%%%%%%%%%%%%%%%%%%%

%
\section{Vortex fermion on the lattice}
In this section we shall put the system of the previous 
section on the lattice. For this purpose, 
we shall first propose the following new 
lattice regularization.

\subsection{New lattice regularization}

\begin{figure}[htb]
\begin{center}
\includegraphics[height=13cm]{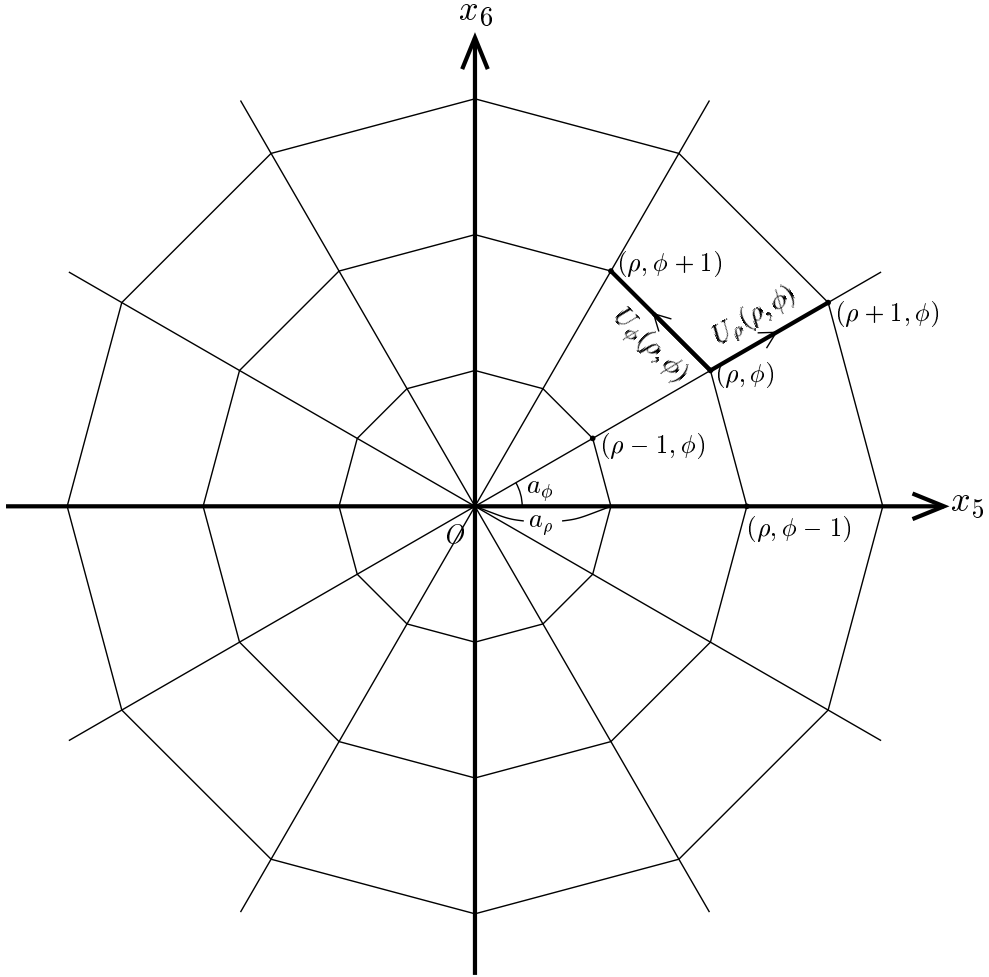}
\end{center}
\caption{The spider's web lattice is drawn.}\label{spider}
\end{figure}

For ${x_1,x_2,\cdots,x_{2n}}$ we discretize 
the Euclidean coordinates as usual, that is, 
square (cubic) lattice which has  
a lattice spacing $a$ for all the directions, while 
for ${x_{2n+1},x_{2n+2}}$ directions we discretize them 
{\it after} taking them as polar coordinates $\rho,\phi$. 
Namely, we now have ``{\it spider's web}'' lattice which has 
lattice spacing $a_\rho$ and ``lattice unit 
angle'' $a_\phi$ corresponding to the polar coordinates 
as in Figure \ref{spider}.\footnote{The lattice spacing 
for the $\phi$ direction 
between the site $(\rho,\phi)$ and $(\rho,\phi+1)$ 
is $2\rho a_\rho \sin{\frac{a_\phi}{2}}$.}
The square lattice which we are familiar with respects 
the discrete translational invariance but sacrifices the rotational 
one and this is always used for constructing various 
models on the lattice\cite{Wilson}. 
But when we want to discretize the model 
which especially respects the rotational invariance rather 
than the translational one, the square lattice is not 
appropriate for describing such a model. 
It hence makes sense that we discretize 
polar coordinates so that the new lattice respects 
the discrete {\it rotational} invariance rather than 
the translational one.
We propose to use 
this new spider's web lattice regularization 
in the case of our vortex fermion model where the 
rotational invariance has an important role.

We define the forward and backward difference 
operator as 
\begin{eqnarray}
\nabla_\rho \psi_{\rho,\phi}&=& 
\frac{\psi_{\rho+1,\phi}-\psi_{\rho,\phi}}{a_\rho},\\
\nabla_\rho^* \psi_{\rho,\phi}&=& 
\frac{\psi_{\rho,\phi}-\psi_{\rho-1,\phi}}{a_\rho},\\
\nabla_\phi \psi_{\rho,\phi}&=& 
\frac{\psi_{\rho,\phi+a_\phi}-\psi_{\rho,\phi}}{2\sin{\frac{a_\phi}{2}}},\\
\nabla_\phi^* \psi_{\rho,\phi}&=& 
\frac{\psi_{\rho,\phi}-\psi_{\rho,\phi-a_\phi}}{2\sin{\frac{a_\phi}{2}}}
\end{eqnarray}
and also the hermitian operator as 
\begin{eqnarray}
\nabla_\rho^h \psi_{\rho,\phi}&=& 
\frac{\nabla_\rho + \nabla_\rho^*}{2}\psi_{\rho,\phi},\\
\nabla_\phi^h \psi_{\rho,\phi}&=& 
\frac{\nabla_\phi + \nabla_\phi^*}{2}\psi_{\rho,\phi}
\end{eqnarray}
though the last one is not used in our present model.

Here we note that $\nabla_\rho^* \psi_{\rho,\phi}$ and 
$\nabla_\rho^h \psi_{\rho,\phi}$ are defined for $\rho\geq 1$ 
and that $\nabla_\phi \psi_{0,\phi}=\nabla_\phi^* 
\psi_{0,\phi}=\nabla_\phi^h \psi_{0,\phi}= 0$.
In this new regularization $\rho$ is defined 
in the region $\rho\geq 0$, so this means that we have 
always a boundary at $\rho=0$ even when we consider 
the infinite volume lattice.\footnote{In this paper 
we consider only the infinite volume lattice.}
Therefore the definition at $\rho=0$ is non-trivial and 
has some ambiguity, which leaves something 
to be discussed. 
In fact if we write the action naively, then 
the action includes a free parameter at $\rho=0$ 
which should be fixed in some appropriate way.
In this paper we impose the constraint of 
$\psi(\rho=1,\phi)=\psi(\rho=0)$ on the action 
and redefine the fixed classical scalar field $\Phi$ 
in order to avoid this ambiguity. 
We note that the $\phi$ dependence of the fields 
at $\rho=1$
is eliminated with this constraint. 
The explicit action satisfying our requirements 
is proposed and discussed in the next subsection.

As for the link variables we define as follows
\begin{eqnarray}
U_\rho(\rho,\phi)&=&
\exp(ig a_\rho A_\rho(\rho,\phi)),\\
U_\phi(\rho,\phi)&=&
\exp(ig 2\rho a_\rho \sin{\frac{a_\phi}{2}}
A_\phi(\rho,\phi)),
\end{eqnarray}
which connect the reference site $(\rho,\phi)$ with the 
neighbor site 
$(\rho+1,\phi)$ and $(\rho,\phi+a_\phi)$  
by gauge fields $A_\rho,A_\phi$ respectively, and 
their backward operators too. 
In this paper we consider only the simplest case 
$U_\rho(\rho,\phi)=U_\phi(\rho,\phi)=1$, that is to say, 
we deal with the dynamical link variables  
only in $4$ dimensional parts ($U_i(x)$ for $i=1,2,3,4$).

Apart from the vortex fermion model in this paper, 
we hope that this 
new lattice regularization will be generally adopted 
for other models where the rotational 
invariance has an important role.

\subsection{Naive construction}
Now we construct the vortex fermion model on the lattice 
using the new lattice 
regularization in extra $2$ dimensions.
As mentioned above, if we write the action naively, 
the action $S_{naive}(b)$ includes a free parameter $b$. 
We refer the explicit form of $S_{naive}(b)$ to 
appendix.
We must adjust the parameter so that we can obtain a desirable 
solution. Zero mode solutions are, in general, the 
relations among three sites $\rho-1,\rho,\rho+1$, 
for various values of $\rho$.
But in order to obtain a {\it unique} zero mode solution, 
this relation must begin from the relation between 
{\it two} sites at the smallest $\rho$, for example, 
the relation between $\rho=0,1$ or $\rho=1,2$ etc.. 
If the zero mode solution begins from the relation 
among three sites, for example, $\rho=0,1,2$ or 
$\rho=1,2,3$ etc., then we cannot determine the solution 
{\it uniquely}, that is to say, 
the solution is not solved inductively 
with the input of an initial 
value at the smallest $\rho$. 
Namely a boundary condition at $\rho=0$ 
must be chosen properly by fixing the free parameter 
in order to ensure the {\it uniqueness} of the solution. 

We find that one of the appropriate methods of parameter fixing 
is to assume the constraint of 
$\psi(\rho=1,\phi)=\psi(\rho=0)$ on the action $S_{naive}(b)$ 
and redefine the fixed classical scalar field $\Phi$. 
This constraint eliminates the terms 
which include the free parameter $b$ in $S_{naive}(b)$, 
rather than fixing the parameter.
As will be shown later, this constraint 
gives a desirable solution, which begins 
from the relation between {\it two} sites 
at the smallest $\rho$.
We also discuss another method of parameter fixing 
of $b$ in appendix. In this subsection we use 
the action $S_{naive}^c$ which is obtained by 
imposing the constraint on $S_{naive}(b)$.

The action $S_{naive}^c$, which does not include the 
free parameter, is written by 
\begin{equation}
  S_{naive}^c=\sum_x \sum_\phi \sum_\rho \rho a_\rho 
\left[\bar\psi \Gamma_\mu \nabla_\mu \psi + 
\bar\psi (\Phi_1 + i\bar\Gamma\Phi_2) 
\psi \right] \label{S_naive^c} ,
\end{equation}
where
\begin{equation}
\sum_\phi \sum_\rho \rho a_\rho \bar\psi 
\Gamma_\mu \nabla_\mu \psi = 
\sum_\phi \sum_\rho \rho a_\rho \bar\psi 
(\Gamma_\mu \nabla_\mu)_{int}\psi 
+\sum_\phi \sum_\rho \rho a_\rho \bar\psi 
(\Gamma_\mu \nabla_\mu)_{ext} \psi \quad ,
\end{equation}
\begin{eqnarray}
&&\sum_\rho \sum_\phi \rho a_\rho 
\bar\psi(\Gamma_\mu \nabla_\mu)_{ext} \psi \nonumber\\
&=& 
\sum_{\rho=1}^{\infty}\sum_{\phi}\rho a_\rho 
\bar\psi_{\rho,\phi}\Gamma_{2n+1}
\left[\left(\cos\phi + i\Gamma_{ext}\sin\phi \right)\nabla_{\rho}^h 
\psi_{\rho,\phi} \right. \nonumber\\
&&-\frac{1}{2\rho a_\rho} 
\left\{  \left(\sin(\phi+\frac{a_\phi}{2})  
-i\Gamma_{ext}\cos(\phi+\frac{a_\phi}{2})\right)
\nabla_\phi \right. \nonumber \\ 
&&\left.+ \left(\sin(\phi-\frac{a_\phi}{2})  
-i\Gamma_{ext}\cos(\phi-\frac{a_\phi}{2})\right)\nabla_\phi^* 
\right\}
\left. \times 
\left(\frac{\psi_{\rho+1,\phi}+\psi_{\rho-1,\phi}}{2} \right)
\right] ,
\end{eqnarray}
and we redefine the fixed classical scalar field $\Phi$ 
as follows
\begin{eqnarray}
\Phi&=&f(\rho)e^{i\varphi}, \\
f(\rho)&=&m_0 \theta(\rho)\equiv 
\frac{\sinh(a_\rho\nu)}{a_\rho} \theta(\rho) \quad,\\
\theta(\rho)&\equiv& \left\{
\begin{array}{ll}
1 & \rho\geq 2 \\
0 & \rho =1 .
\end{array}
\right.  
\end{eqnarray}

%%%%%%%%%%%%%%%%%%%%%%%%%%%%%%%%%%%%%%%%%%%%%%%%%%%%%%%%%%%%%

Now we seek the zero mode solution.
In order to obtain the Dirac equation, we vary the 
action $S_{naive}^c$ by $\bar\psi_{\rho,\phi}$ 
according to the value of $\rho$.
The variation is as follows:
\begin{enumerate}
%%%%%
\item by $\frac{\delta}{\delta\bar\psi_{1}}$ \\
%%%%%
The Dirac equation becomes 
\begin{eqnarray}
&&a_\rho\fslash{\nabla}_{int}\psi_{1,\phi}
+\sum_\phi a_\rho \Gamma_{2n+1}
 \left[ \right.     
e^{i\Gamma_{ext}\phi}\nabla_{\rho}^h 
\psi_{\rho,\phi}\vert_{\rho=1}  \nonumber\\
&&+\frac{i}{2a_\rho} \Gamma_{ext}
\left\{  e^{i\Gamma_{ext}(\phi+\frac{a_\phi}{2})}
\nabla_\phi 
+e^{i\Gamma_{ext}(\phi-\frac{a_\phi}{2})}
\nabla_\phi^*\right\} 
\left. \times 
\left(\frac{\psi_{2,\phi}+\psi_{1}}{2} \right)
\right] \nonumber\\
&&=0,
\end{eqnarray}
which is equivalent to 
\begin{eqnarray}
&&\fslash{\nabla}_{int}\psi_{1}^{\beta}
+
\sum_\phi \Gamma_{2n+1}
\left[
e^{i\Gamma_{ext}\phi}
\nabla_\rho^h \psi_{\rho,\phi}\vert_{\rho=1}^{\beta} 
-\frac{1}{2a_\rho}
(-i\Gamma_{ext})\times
\right.\nonumber \\
&&
\left.
\left\{
e^{i\Gamma_{ext}(\phi+\frac{a_\phi}{2})}\nabla_\phi
+e^{i\Gamma_{ext}(\phi-\frac{a_\phi}{2})}\nabla_\phi^*
\right\}
\times
\left(\frac{\psi_{2,\phi}^\beta+\psi_{1}^\beta}{2} \right)
\right]
=0, \\
&&\fslash{\nabla}_{int}\psi_{1}^{\alpha}
+
\sum_\phi \Gamma_{2n+1}
\left[
e^{i\Gamma_{ext}\phi}
\nabla_\rho^h \psi_{\rho,\phi}\vert_{\rho=1}^{\alpha} 
-\frac{1}{2a_\rho}
(-i\Gamma_{ext})\times
\right.\nonumber \\
&&
\left.
\left\{
e^{i\Gamma_{ext}(\phi+\frac{a_\phi}{2})}\nabla_\phi
+e^{i\Gamma_{ext}(\phi-\frac{a_\phi}{2})}\nabla_\phi^*
\right\}
\times
\left(\frac{\psi_{2,\phi}^\alpha+\psi_{1}^\alpha}{2} \right)
\right]
=0.
\end{eqnarray}
We substitute the following form into the zero mode equation 
\begin{eqnarray}
\psi_\beta^{\pm} &=&
\left\{
\begin{array}{cl}
&e^{ip\cdot x}
\varphi_{\pm}(\rho,\mbox{\boldmath $p$},k) e^{ik\phi} u_{\pm} 
\quad \mbox{for}\quad \rho \geq 2, \\
&e^{ip\cdot x}
\varphi_{\pm}(\rho,\mbox{\boldmath $p$})u_{\pm}
 \quad \mbox{for}\quad \rho =1,
\end{array}
\right.
\\
\psi_\alpha^{\pm} &=& \Gamma_{2n+1}\psi_\beta^{\pm}.
\end{eqnarray}
Here we note that the value of $k$ is restricted to $0,1,2,\cdots, N_\phi-1$, 
where $N_\phi$ is the number of lattice sites in the angular direction, 
namely, $N_\phi=\frac{2\pi}{a_\phi}$.
We keep in mind this restriction on $k$ whenever we deal with the spider's web lattice.
Then the zero mode equation becomes  
\begin{eqnarray}
\varphi_{\pm}(2)-\varphi_{\pm}(1)e^{-ik\phi}
\mp \frac{i \varphi_{\pm}(2)}{4\sin{ \frac{a_\phi}{2} }}
%\cdot 
\left\{
e^{\mp i \frac{a_\phi}{2}}\left( e^{ika_\phi}-1  \right)
+
e^{\pm i \frac{a_\phi}{2}}\left( 1-e^{-ika_\phi}  \right)
\right\}
=0,&&\\
\varphi_{\pm}(2)-\varphi_{\pm}(1)e^{-ik\phi}
\pm \frac{i \varphi_{\pm}(2)}{4\sin{\frac{a_\phi}{2}}}
%\cdot 
\left\{
e^{\pm i \frac{a_\phi}{2}}\left( e^{ika_\phi}-1  \right)
+
e^{\mp i \frac{a_\phi}{2}}\left( 1-e^{-ika_\phi}  \right)
\right\}
=0.&&
\end{eqnarray}
%%%%%%%%%%
%
A necessary condition for the existence of the solution is 
to choose $k$ such that $e^{ik\phi}=1$.
%We need to choose $k$ such that $e^{ik\phi}=1$.
Generally, $k=\frac{n\pi}{a_\phi} \ {} (n \in \mbox{\boldmath $Z$}_{even})$ satisfies 
$e^{ik\phi}=1$, but we have the restriction on $k$ mentioned above.
%
%%%%%%
Thus the solution exists for $k=0$, and it is given by
%$k=\frac{n\pi}{a_\phi} {} (n \in \mbox{\boldmath $Z$}_{even})$, 
%in which case $e^{ik\phi}=1$, and it is given by 
\begin{equation}
\varphi_{\pm}(2)=\varphi_{\pm}(1).
\end{equation}
%

%%%%%%%%%%
\item by $\frac{\delta}{\delta\bar\psi_{\rho,\phi}}$ for $\rho\geq 2$\\
%%%%%%%%%%
%
Solving the zero mode equation as the case above, 
we obtain the minus chirality solution
\begin{equation}
\varphi_-(\rho+1)-\varphi_-(\rho-1)=-2a_\rho 
f(\rho)\varphi_-(\rho).
\end{equation}

\end{enumerate}

From the consideration above, we have obtained 
the following normalizable solution for $k=0$
\begin{equation}
\left\{
\begin{array}{cl}
\varphi_-(2)&=\varphi_-(1),  \\
\varphi_-(\rho+1)&=\varphi_-(\rho-1)-2a_\rho 
f(\rho)\varphi_-(\rho) \quad \mbox{for}\quad \rho \geq 2 .
\end{array}
\right.
\end{equation}
%for $k=\frac{n\pi}{a_\phi} {} (n \in \mbox{\boldmath $Z$}_{even})$, 
%in which case $e^{ik\phi}=1$.
%
This solution begins from the relation 
between two sites at the smallest $\rho$, so 
we can obtain a unique $\varphi_-(\rho)$ from $\rho=2$ to 
$\rho=\infty$ inductively with the input of $\varphi_-(1)$. 

%%%%%%%%%%%%%%%%%%%%%%%%%%%%%%%%%%%%%%%%%%%%%%%%%%%%%%%%%%%%%%%%%%%%%%

But $\fslash{\nabla}_{int}\psi$ is written as 
\begin{equation}
\sum_{i=1}^{2n}\frac{i}{a}\Gamma_{i}\sin(p_i a)
\psi(\mbox{\boldmath $p$},z)
\end{equation}
in momentum space, so doubling species appear.

Here we compare the appearance of doublers in 
vortex fermion with that in the case of domain wall fermion.
In the latter case, {\it both} chirality 
solutions localized at the defect 
appear for $\mbox{\boldmath $p$}=(0,0,0,0)$ in $4$ dimensional 
momentum space and 
from the term $\sin(p_ia)$
there appear the flipped 
chirality solutions for 
$\mbox{\boldmath $p$}=(\frac{\pi}{a},0,0,0)$ etc. 
corresponding to each chirality solution.
On the other hand, in our vortex fermion, 
only a {\it single} (minus) chirality solution localized 
at $\rho=0$ appears for $\mbox{\boldmath $p$}=(0,0,0,0)$, 
because $e^{i\bar\Gamma\phi}$ factor determines 
a single chirality 
definitely.\footnote{If we replace the mass term 
$\Phi_1 + i\bar\Gamma \Phi_2$ to 
$\Phi_1 - i\bar\Gamma \Phi_2$ in the action, 
we obtain an opposite (positive) chirality solution. 
This is an anti-vortex fermion which appears 
at a large distance when the volume is finite. 
This situation looks like a domain wall fermion case. }
This is quite a 
different point from the case of domain wall fermion.
Though there appear the flipped 
chirality solutions at 
$\mbox{\boldmath $p$}=(\frac{\pi}{a},0,0,0)$ etc. also 
in this vortex fermion from the term $\sin(p_i a)$, 
the number of doubling species is reduced to almost half 
of that in the case of domain wall fermion.
This better point of vortex fermion than that of domain 
wall fermion is owing to the fact that vortex fermion has 
one more extra-dimension than domain wall fermion and 
that one is therefore allowed to have a little more 
freedom to give more adjusted scalar fields or mass terms 
to the model in order to obtain a desired solution 
which has better properties. 

At any rate, this naive construction model has 
doubling species, 
so we shall consider using Wilson term in the next subsection 
in order to eliminate the doubling species.

\subsection{Extended Wilson term}
If we use the usual Wilson term\cite{Wilsonterm}, 
where extra $2$ dimensions in Euclidean coordinates are 
taken as polar coordinates, 
we cannot obtain zero mode solutions with a definite 
chirality.
This is because the usual Wilson term does not include the 
$e^{i\bar\Gamma\phi}$ factor 
which has an important role of determining the 
chirality of the zero mode solutions definitely.
Therefore we need to extend 
the usual Wilson term so that we can still keep 
the $e^{i\bar\Gamma\phi}$ factor even 
after we add the extended 
one to $S_{naive}^c$. 
Furthermore when we construct naively the extended Wilson term 
$S_W(c)$ which has the hermiticity in both $\rho$ and $\phi$ 
directions, $S_W(c)$ includes a free parameter $c$ like 
$b$ in the naive action $S_{naive}(b)$.
We refer the explicit form of the extended Wilson term 
$S_W(c)$ to appendix.
Since $S_W(c)$ has some ambiguity at the $\rho=0$ boundary, 
we have to adjust this parameter to obtain a desirable 
solution.

We now impose the same constraint as in previous subsection, 
$\psi(\rho=1,\phi)=\psi(\rho=0)$ on $S_W(c)$ and use the 
redefined scalar field $\Phi$.
Then this constraint eliminates the terms which 
include the free parameter $c$ in $S_W(c)$, 
and gives a desirable solution as shown later.
In appendix we also discuss 
another method of parameter fixing of $b,c$ in 
the whole action $S_{naive}(b)+S_W(c)$.

We now use the action $S_W^c$ which is obtained by 
imposing the constraint on the action $S_W(c)$.
The action $S_W^c$, which does not include a free 
parameter $c$, is written by 
\begin{eqnarray}
S_W^c &=& S_W^{c(in)} + S_W^{c(out)} ,\label{S_W^c}\\
S_W^{c(in)}&=&\sum_x \sum_{\rho=1}^{\infty}\sum_{\phi}
\frac{wa}{m_0} 
\sum_{i=1}^{2n} \bar\psi_{\rho,\phi} \rho a_\rho
(\Phi_1 +i\bar\Gamma\Phi_2)
\Delta_i \psi_{\rho,\phi} ,\\
S_W^{c(out)}&=&\sum_x \sum_{\rho=1}^{\infty}
\sum_{\phi}
\frac{w_{\rho} a_{\rho}}{m_0}
\bar\psi_{\rho,\phi} \rho a_\rho
\left[(\Phi_1 +i\bar\Gamma\Phi_2)
\Delta_\rho \psi_{\rho,\phi}
+\Delta_\rho \left((\Phi_1 +i\bar\Gamma\Phi_2)\psi_{\rho,\phi}\right)
\right]\times\frac{1}{2} \nonumber \\
&+&\sum_x \sum_{\rho=1}^{\infty}\sum_{\phi}
\frac{w_{\rho} a_{\rho}}{m_0}
\bar\psi_{\rho,\phi} 
\left[(\Phi_1 +i\bar\Gamma\Phi_2)
\nabla_\rho^h \psi_{\rho,\phi}
+\nabla_\rho^h \left((\Phi_1 +i\bar\Gamma\Phi_2)\psi_{\rho,\phi}\right)
\right]\times\frac{1}{2} \nonumber \\
&+&\sum_x \sum_{\rho=1}^{\infty}\sum_{\phi}
\frac{w_\phi a_\phi^2 a_\rho}{m_0}
\bar\psi_{\rho,\phi}\frac{1}{\rho a_\rho}
\left[(\Phi_1 +i\bar\Gamma\Phi_2)
\Delta_\phi \psi_{\rho,\phi}
+\Delta_\phi \left((\Phi_1 +i\bar\Gamma\Phi_2)\psi_{\rho,\phi}\right)
\right]\times\frac{1}{2}.  \nonumber\\
\end{eqnarray}
Using this extended Wilson term $S_W^c$ 
we construct vortex fermion 
model in the next subsection.

\subsection{Simple vortex model}

We first construct a simple vortex model from which 
we can see easily that there indeed exists a normalizable 
zero mode solution localized at the string.

The action is 
\begin{equation}
S=S_{naive}^{c,\nabla_\rho^h \rightarrow \nabla_\rho} 
+ S_W^{c(in)}
\end{equation}
where 
$S_{naive}^{c,\nabla_\rho^h \rightarrow \nabla_\rho}$ 
is obtained by replacing $\nabla_\rho^h$ to $\nabla_\rho$ 
in $S_{naive}^c$.
Here we sacrificed the 
hermiticity in the $\rho$ direction at a finite 
lattice spacing $a_\rho$ in order to see clearly that we 
have a normalizable zero mode solution.

%%%%%%%%%%%%%%%%%%%%%%%%%%%%%%%%%%%%%%%%%%%%%%%%%%%%%%%%%%%%%%%%%%%%%%%%%%%%%

Now we seek the zero mode solution.
In order to obtain the Dirac equation, we vary the action 
by $\bar\psi_{\rho,\phi}$ according to the value of
$\rho$.
The variation is as follows: 
\begin{enumerate}
%%%%%%%
\item by $\frac{\delta}{\delta\bar\psi_{1}}$ \\
%%%%%%%
%
The Dirac equation becomes 
\begin{eqnarray}
&&a_\rho \fslash{\nabla}_{int}\psi_{1}
+\sum_\phi a_\rho\Gamma_{2n+1}
\left[ \right.     e^{i\Gamma_{ext}\phi}
\nabla_{\rho} \psi_{\rho,\phi} \vert_{\rho=1}  \nonumber \\
&&+\frac{i}{2a_\rho} \Gamma_{ext}
\left\{  
e^{i\Gamma_{ext}(\phi+\frac{a_\phi}{2})}
\nabla_\phi 
+
e^{i\Gamma_{ext}(\phi-\frac{a_\phi}{2})}
\nabla_\phi^*\right\}
\left.\times 
\left(\frac{\psi_{2,\phi}+\psi_{1}}{2} \right)
\right]  \nonumber \\
&=&0
\end{eqnarray}

Solving the zero mode equation as in the previous subsection, 
we obtain 
\begin{equation}
\varphi_{\pm}(2)=\varphi_{\pm}(1). 
\end{equation}
This solution exists for 
$k=0$.
%$k=\frac{n\pi}{a_\phi} {} (n \in \mbox{\boldmath $Z$}_{even})$, 
%in which case $e^{ik\phi}=1$. 

%%%%%%%%%%
\item by $\frac{\delta}{\delta\bar\psi_{\rho,\phi}}$ 
for $\rho\geq 2$\\
%%%%%%%%%%
%
The Dirac equation becomes 
\begin{eqnarray}
&& \fslash{\nabla}_{int}\psi_{\rho,\phi}
+\Gamma_{2n+1}
\left[ \right.     
e^{i\Gamma_{ext}\phi}
\nabla_{\rho} 
\psi_{\rho,\phi}  \nonumber \\
&&+\frac{i}{2\rho a_\rho}\Gamma_{ext} \left\{  e^{i\Gamma_{ext}(\phi+\frac{a_\phi}{2})}
\nabla_\phi 
+ 
e^{i\Gamma_{ext}(\phi-\frac{a_\phi}{2})}
\nabla_\phi^*\right\}
\left.\times 
\left(\frac{\psi_{\rho+1,\phi}+\psi_{\rho-1,\phi}}{2} \right)
\right]  \nonumber \\
&=&
- f(\rho)
e^{i\bar\Gamma \phi}
\psi_{\rho,\phi} 
-\frac{wa}{m_0} \sum_{i=1}^{2n} 
f(\rho)e^{i\bar\Gamma \phi}
\Delta_i \psi_{\rho,\phi}. 
\end{eqnarray}
Solving the zero mode equation, 
we obtain the minus chirality solution
\begin{equation}
\varphi_-(\rho+1)=z\varphi_-(\rho),
\end{equation}
where we have defined $F(\mbox{\boldmath $p$})\equiv
\sum_{i =1}^{2n}(1-\cos{ap_i}) \geq 0, z=1-a_\rho m_0 +F$ and 
chosen $w$ as $w=\frac{a}{2a_\rho}.$

\end{enumerate}
%

%%%%%%%%%%%%%%%%%%%%%%%%%%%%%%%%%%%%%%%%%%%%%%%%%%%%%%%%%%%%%%%%%%%%%%%%%%%

From the consideration above we obtain for $k=0$ 
\begin{eqnarray}
\varphi_-(2)&=&\varphi_-(1), \nonumber \\
\varphi_-(\rho+1)&=&z\varphi_-(\rho) \quad (\rho \geq 2) .
\end{eqnarray}
We see that 
$\varphi_-(\rho)$ is normalizable if 
$\vert z \vert <1  \Leftrightarrow 0<a_\rho m_0 -F(\mbox{\boldmath $p$})<2$ 
is satisfied.
Noting that $F(0)=0$, a normalizable solution exists 
for small $\mbox{\boldmath $p$}$ only if $0<a_\rho m_0<2$.
We also see that  
for $F\geq 2$, which means the appearance of doubling species, 
there is no normalizable solution
because $0<a_\rho m_0<2$ is not satisfied.\footnote{We note 
here that we can exclude doubling species by 
the normalizable condition with Wilson term.}
This situation looks like that in the case 
of domain wall fermion.\footnote{In domain wall fermion case 
it is shown that a chiral zero mode solution exists 
for wide range value of $w$\cite{wide}.}

In this subsection we have seen that there exists 
a normalizable 
zero mode solution localized at the string in the simple 
vortex model, just as in the well-known continuum case.

\subsection{Hermitian vortex model}

In this subsection we shall construct the hermitian vortex model 
and show that there exits a normalizable zero mode solution 
localized at the string.
 
The action is defined as 
\begin{equation}
S=S_{naive}^c+S_W^c. \label{para_bc}
\end{equation}
We seek the zero mode solution. 
In order to obtain the Dirac equation, we vary the action 
by $\bar\psi_{\rho,\phi}$ according to the value of
$\rho$.
The variation is as follows:
\begin{enumerate}
%%%%%%%%%%%
\item by $\frac{\delta}{\delta\bar\psi_{1,\phi}}$ \\
%%%%%%%%%%%
The Dirac equation becomes 
\begin{eqnarray}
&&a_\rho \fslash{\nabla}_{int}\psi_{1}
+\sum_\phi a_\rho\Gamma_{2n+1}
\left[ \right.     
e^{i\Gamma_{ext}(\phi)}\nabla_{\rho}^h 
\psi_{\rho,\phi}\vert_{\rho=1}  \nonumber \\
&&+\frac{i}{2a_\rho} \Gamma_{ext} 
\left\{  e^{i\Gamma_{ext}(\phi+\frac{a_\phi}{2})} \nabla_\phi 
+ 
e^{i\Gamma_{ext}(\phi-\frac{a_\phi}{2})}
\nabla_\phi^*\right\}
\left. \times 
\left(\frac{\psi_{2,\phi}+\psi_{1}}{2} \right)
\right]  \nonumber \\
&=&
-\sum_\phi \frac{w_{\rho}a_{\rho}^2}{m_0} 
\left[
\Delta_\rho \left\{(\Phi_1 +i\bar\Gamma\Phi_2)
\psi_{\rho,\phi}\right\} \vert_{\rho=1}
\right]
\times\frac{1}{2} \nonumber \\
&&-\sum_\phi \frac{w_{\rho}a_{\rho}}{m_0} 
\left[
\nabla_\rho^h \left\{ (\Phi_1 +i\bar\Gamma\Phi_2)
\psi_{\rho,\phi} \right\}\vert_{\rho=1}
\right]
\times\frac{1}{2}. 
\end{eqnarray}
Solving the zero mode equation as in the previous subsection, 
we obtain 
\begin{eqnarray}
&& \varphi_{-}(2)-\varphi_{-}(1)  e^{-ik\phi} 
+\frac{i}{4 \sin{\frac{a_\phi}{2}}} 
\left\{ e^{i \frac{a_\phi}{2}}(e^{ik a_\phi}-1)
+e^{-i \frac{a_\phi}{2}}(1-e^{-ik a_\phi})
\right\}
\varphi_-(2) \nonumber \\
&=&
-\frac{3w_{\rho}}{2} \varphi_{-}(2) , \\
&& \varphi_{-}(2)-\varphi_{-}(1)  e^{-ik\phi} 
-\frac{i}{4 \sin{\frac{a_\phi}{2}}} 
\left\{ e^{-i \frac{a_\phi}{2}}(e^{ik a_\phi}-1)
+e^{i \frac{a_\phi}{2}}(1-e^{-ik a_\phi})
\right\}
\varphi_-(2) \nonumber \\
&=&
-\frac{3w_{\rho}}{2} \varphi_{-}(2) . 
\end{eqnarray}

We thus obtain the minus chirality solution for $k=0$ 
\begin{equation}
\varphi_-(2)=\frac{2}{2+3 w_\rho}\varphi_-(1)  .
\end{equation}

\item by $\frac{\delta}{\delta\bar\psi_{\rho,\phi}}$ for 
$\rho=2$\\
Solving the zero mode equation as the case above, 
we obtain 
\begin{equation}
\varphi_-(3)=
\frac{
\left[4w_\rho + \frac{w_\phi a_\phi^2}{4} -2a_\rho m_0 +4wF \frac{a_\rho}{a}\right]
\varphi_-(2)
+ \left( 1-\frac{3}{4} w_\rho \right) \varphi_-(1)}
{1+\frac{5}{2}w_\rho}. 
\end{equation}

\item by $\frac{\delta}{\delta\bar\psi_{\rho,\phi}}$ for 
$\rho\geq 3$\\
Solving the zero mode equation in the same way, 
we obtain 
\begin{equation}
\varphi_-(\rho+1)=
\frac{
\left[2w_\rho-a_\rho m_0 + \frac{2w a_\rho F}{a} 
+ \frac{w_\phi a_\phi^2}{2\rho^2} \right]
\varphi_-(\rho)
+\left( \frac{1}{2}-w_\rho + \frac{w_\rho}{2\rho} \right) \varphi_-(\rho-1)}
{\frac{1}{2}+w_\rho + \frac{w_\rho}{2\rho}}.          
\end{equation}
\end{enumerate}
We choose $w_\rho$, $w_\phi$ and $w$ such that $w_\rho=w_\phi a_\phi^2=\frac{1}{2}$ and 
$\frac{w a_\rho}{a}=\frac{1}{2}$.
Then the above solution, which exists for $k=0$, is expressed as follows:
\begin{eqnarray}
&&\varphi_-(2)=\frac{4}{7}\varphi_-(1), \label{phi_2a} \\
&&\varphi_-(3)=
\frac{
\left[ 1-a_\rho m_0 \left( 1-\frac{1}{16 a_\rho m_0} \right) +F \right]
\varphi_-(2)
+ \frac{5}{16} \varphi_-(1)}{ \frac{9}{8} }, \label{phi_rho2} \\
&&\varphi_-(\rho+1)=
\frac{\left[1-a_\rho m_0\left(1-\frac{1}
{4\rho^2 a_\rho m_0}\right) +F \right]\varphi_-(\rho)
+\frac{1}{4\rho}\varphi_-(\rho-1)}
{1 + \frac{1}{4\rho}}
\quad {\text{for}} \quad \rho \geq  3.  \nonumber   \\ \label{phi_rho3}
\end{eqnarray}

We note here that the solution begins from the relation 
between {\it two} sites at the smallest $\rho$ 
as eq.(\ref{phi_2a}), 
and that if we input $\varphi_-(1)$ 
into eqs.(\ref{phi_2a})(\ref{phi_rho2})(\ref{phi_rho3}), 
then we obtain a {\it unique} $\varphi_-(\rho)$ 
from $\rho=2$ to $\rho=\infty$ 
inductively.
If the parameters $b,c$ in the action $S_{naive}(b)+S_W(c)$ 
are not adjusted properly, 
the solution at the smallest $\rho$, corresponding to 
eq.(\ref{phi_2a}), becomes the relation between three sites, 
$\rho=0,1,2$ in general, which means that we cannot 
determine the solution of $\varphi_-(\rho)$ uniquely 
with the input of $\varphi_-(0)$.
Our constraint of $\psi(\rho=1,\phi)=\psi(\rho=0)$ 
on $S_{naive}(b)+S_W(c)$ and redefinition of $\Phi$, 
a kind of the parameter fixing, avoids this problem 
and works well.

Since we have solved the zero mode equations, we now 
consider the normalizable condition.
We define $P_\rho$ as 
\begin{equation}
\varphi_-(\rho+1)=P_\rho \varphi_-(\rho) \label{p_rho},
\end{equation}
then we find that 
for large $\rho$  
\begin{equation}
P_\rho \simeq 1-a_\rho m_0 +F = z \label{p_rhoz}.
\end{equation}
We see that the $\varphi_-(\rho)$ is normalizable if 
$\vert z \vert <1$ is satisfied.
Noting that $F(0)=0$, a normalizable zero mode solution exists 
for small $\mbox{\boldmath $p$}$ only if $0<a_\rho m_0<2$.
Furthermore we see that  
for $F\geq 2$, which means the existence of species 
doublers, 
there is no normalizable zero mode solution
because $0<a_\rho m_0<2$ is not satisfied.
We have thus shown that a normalizable zero mode solution 
localized at the string exists also in this 
hermitian vortex model.

%%%%%%%%%%%%%%%%%%%%%%%%%%%%%%%%%%%%%%%%%%%%%%%%%%%%%%%%%%%%%%%%%%%

%%%%%%%%%%%%%%%%%%%%%%%%%%%%%%%%%%%%%%%%%%%%%%%%%%%%%%%%%%%%%%%%%%%

Finally we briefly mention the subtraction 
of massive modes, though this is not a real analysis.
Our model $D_{\text{vortex}}$ includes many massive modes besides the 
zero mode solution localized at the string.
Since we want only the zero mode solution, we need to 
construct the Pauli-Villars Dirac operator 
$D_{\text{vortex}}^{\text{PV}}$ 
in order to subtract the unwanted modes.
In the domain wall fermion case the absence of 
Pauli-Villars zero modes is obvious once anti-periodic 
boundary conditions are used. 
In our spider's web lattice, 
considering the situation that the zero mode solution is localized 
at the $\rho=1$ string, we assume the Pauli-Villars 
Dirac operator $D_{\text{vortex}}^{\text{PV}}$ which has a 
constraint of $\psi(\rho=1)=0$. 
This constraint is expected to exclude only the zero 
mode solution.
Then the interface Dirac operator $D_{\text{vortex}}^i$ 
is written as  
\begin{eqnarray}
\det{D_{\text{vortex}}^i} &=&
\lim_{{\scriptstyle a_\rho,a_\phi\to 0} \atop 
{\scriptstyle N_\rho \to \infty}}
\frac{\det{D_{\text{vortex}}}}
{\det{D_{\text{vortex}}^{\text{PV}}}} \nonumber\\
&=&
\lim_{{\scriptstyle a_\rho,a_\phi\to 0} \atop 
{\scriptstyle N_\rho \to \infty}}
\frac{\int d\psi d\bar\psi e^{-S}}
{\int d\psi d\bar\psi 
\delta\left(\psi(x,1,\phi)\right)e^{-S}}. \label{subtraction}
\end{eqnarray}
where we note that $a_\phi N_\phi =2\pi$.
This 
$\det{D_{\text{vortex}}^i}$\footnote{This 
$\det{D_{\text{vortex}}^i}$ is expected to correspond to 
the chiral determinant of overlap formula 
which was derived by Narayanan and Neuberger 
from domain wall fermion in 
$5$ dimensions\cite{twovacuua}\cite{construction}.} 
is expected to 
include only the zero mode 
solution localized 
at the string, and produce the correct anomaly.
Besides this direct calculation, 
we expect that  
the $2n$ dimensional anomaly 
is observed as the charge flow of $2n+2$ dimensional 
space off the string in our model.
It is important and interesting to check if 
the correct anomaly is produced in both ways 
on the lattice\cite{nagao}.

\section{Conclusions and discussions}
In this paper we constructed a chiral vortex fermion model 
on the lattice.
In section $2$ we first studied vortex fermion in the 
continuum theory. 
In section $3$ we proposed a new lattice 
regularization, spider's web lattice, which 
has the discrete rotational symmetry rather than the 
translational one.
We obtained this by discretizing polar coordinates. 
We hope that this spider's web lattice is 
useful to invesigate some models where 
the rotational symmetry has an important role.

The action constructed on the spider's web 
lattice includes 
a free parameter which should be fixed properly.
We proposed the reasonable constraints so that 
we can obtain a desirable solution avoiding this ambiguity.
We also discussed another example of parameter fixing 
in appendix. 
Next in order to eliminate the doubling species 
which appeared due to the naive construction, 
we introduced the extended Wilson term which was modified 
to include scalar fields appropriate to our model. 
We constructed two models for convenience and 
obtained a normalizable zero mode solution 
localized at the string. 

%
%%%%%%%
In order to make this model a more realistic 
one which we can put on computers, 
we need to construct a vector-like 
vortex and anti-vortex fermion system in finite volume.
The framework is that the string penetrates 
a sphere $S^2$ whose radius is $r$ where 
the vortex fermion appears at the north pole, while 
the anti-vortex fermion at the south pole.
This is a vector-like vortex fermion model.
If we set the radius $r$ to $\infty$, then 
we obtain a chiral vortex model, which we constructed in 
this paper, from a northern hemisphere, 
and a chiral anti-vortex model from a southern hemisphere.
It will be possible to construct the vector-like 
vortex fermion model on the lattice, 
though it will become a little complicated to define 
a ``{\it mirror ball}'' like lattice by 
discretizing the $S^2$ surface.
It might be advantageous for this purpose to regard 
the sphere as a cylinder where the both edges are tied.
%%%%%%%%%%%%%%%%%%%%%%%%%%%%%%%%%%%

If we start from such a vector-like model, 
we should consider the subtraction of 
the unwanted modes and check whether a new Dirac 
operator which will be derived satisfies G-W relation or not.
$D_{\text{vortex}}$ includes many massive modes besides 
the zero mode solution localized at the string.
Since we want only the zero mode solution, we need to 
construct the Pauli-Villars Dirac operator 
$D_{\text{vortex}}^{\text{PV}}$ 
in order to subtract the unwanted modes.
If we can construct $D_{\text{vortex}}^{\text{PV}}$, 
we should perform the following procedure
and check whether 
the following Dirac operator $D_{\text{vortex}}^{\text{GW?}}$ 
satisfies G-W relation or not, 
\begin{equation}
\lim_{{\scriptstyle a_\rho,a_\phi\to 0} \atop 
{\scriptstyle N_\rho \to \infty}}
\frac{\det{D_{\text{vortex}}}}
{\det{D_{\text{vortex}}^{\text{PV}}}}
=\det{D_{\text{vortex}}^{\text{GW?}}}.
\end{equation}
It is natural to expect this occurs, 
because in the case of domain wall 
fermion this is indeed the case.

It has been known that every Dirac operator whose form is 
$D=1+\gamma_5 X$ where $X^2=1$ satisfies 
G-W relation regardless of the 
content of $X$\cite{GWover}, and 
overlap Dirac operator has such a form.
This is, however, a non-trivial thing.
It has not been given any satisfactory explanation 
why overlap Dirac operator 
originating from the idea of extra-dimension 
links to G-W relation which is derived from 
a quite different idea of renormalization group.
Therefore also in this sense it is interesting 
to check if the Dirac 
operator $D_{\text{vortex}}^{\text{GW?}}$ 
satisfies G-W relation or not.
If it does, 
we should investigate the various properties such as locality, 
chirality etc., which may offer another formalism of 
chiral fermion besides the usual overlap Dirac operator.

Finally in order to construct chiral gauge theories 
we must consider 
gauge fields which depend in some manner on the 
extra coordinates. The first of these attempts was the 
so-called waveguide model 
which did not work\cite{Kaplan2}\cite{waveguide}. 
At present a far more sophisticated construction was 
given by Kikukawa which is closely related to L\"uscher's 
construction\cite{kikudomain}. 
To extend this with our vortex fermion, 
we need to start from the vector-like model 
in finite volume.
Though in this paper we have not constructed 
a vector-like model, but a chiral one in infinite volume, 
the work of this paper is a preliminary step to 
proceed along such directions.
Furthermore it may be another direction 
to construct a mechanism where $6$ dimensional gauge fields 
localize into $4$ dimensions.
Indeed in the recent approaches of 
brane world scenario\cite{Ransun}, 
where we think we live in a localized space embedded 
in higher dimensions, various gauge field localization 
mechanisms are proposed in $5$ or $6$ 
dimensions\cite{Oda}. 
Though they are not directly applicable to our lattice model, 
it will be possible to construct similar mechanisms 
also on the lattice.
If we can have such a construction on the lattice, 
it will mean that the interpolation path of gauge fields 
from higher to $4$ dimensional lattice 
does not depend on 
the detail of $5$ and $6$ dimensional gauge field space. 
It is interesting to combine vortex 
fermion with this mechanism for obtaining the 
non-abelian chiral gauge theory on the lattice.
If it is done, the current $j_\mu$ in L\"uscher's 
construction is expected to be determined nonperturbatively 
on the lattice.
This is hopefully another approach 
to the construction of non-abelian chiral gauge theory 
on the lattice 
besides L\"uscher's one where we need the classification 
of the structure of $4+2$ dimensional gauge field space.

\bigskip
\noindent
{\bf Acknowledgments}  \\
The author would like to thank Prof.~T.~Yoneya for many 
useful discussions and encouragement.
He is also grateful to Prof.~Y.~Kikukawa for useful discussions 
and Prof.~T.~Onogi, Prof.~S.~Ichinose, Prof.~H.~So and Mr.~N.~Ukita 
for helpful comments.
Furthermore, on revising this paper, he would like to 
thank Prof.~T.~Yoneya again for valuable discussions, 
and Prof.~K.~Fujikawa and Prof.~S.~Iso for useful comments. 
He also would like to thank 
the author of ref.\cite{Neuberger:2003yg}, 
Prof.~Herbert Neuberger, for giving him the valuable opportunity 
to notice the errors and reconsider the work.
The work of the author is supported in part by Grant-in-Aid for Scientific 
Research No.18740127 from the Ministry of Education, Culture, Sports, 
Science and Technology. 

%%%%%%%%
%%%%%%%

\appendix

\section{Notation of gamma matrices}\label{notation}
We give the explicit notation of the gamma matrices 
which appear in this paper as follows,  
\begin{eqnarray}
\Gamma_i &=&
\left(
\begin{array}{cc}
& -i\gamma_i \\ 
-i\gamma_i &
\end{array}
\right) \quad (i=1,2,\cdots, 2n-1) ,\\
\Gamma_{2n} &=&
\left(
\begin{array}{cc}
& \gamma_{2n} \\ 
\gamma_{2n} &
\end{array}
\right) , \\
\Gamma_{2n+1} &=&
\left(
\begin{array}{cc}
& -i \\ 
i &
\end{array}
\right) , \\
\Gamma_{2n+2} &=&
\left(
\begin{array}{cc}
& -\gamma_{2n+1} \\ 
-\gamma_{2n+1} &
\end{array}
\right) , \\
\gamma_{2n+1}&=&i^{n-1}\gamma_{2n}
\gamma_1\cdots\gamma_{2n-1},
\end{eqnarray}
where
$\Gamma_\mu$ ($\mu=1,2,\cdots,2n+2$) and 
$\gamma_i$ ($i=1,2,\cdots,2n$) are 
$2n+2$ dimensional and $2n$ dimensional gamma matrices 
respectively and $\gamma_{2n+1}$ is the chiral operator 
in $2n$ dimensions.
They satisfy the following hermitian 
and anti-hermitian relations
\begin{eqnarray}
(\gamma_{2n})^\dag &=& \gamma_{2n} , \\
(\gamma_i)^\dag &=& -\gamma_i \quad (i=1,2,\cdots 2n-1),\\
(\gamma_{2n+1})^\dag &=& \gamma_{2n+1}, \\
(\Gamma_\mu)^\dag &=& \Gamma_\mu \quad (\mu=1,2,\cdots 2n+2), \\
(\Gamma_\mu)^2 &=& 1 \quad (\mu=1,2,\cdots 2n+2).
\end{eqnarray}

Using the above gamma matrices we define 
the $2n+2$ dimensional chiral operator $\bar\Gamma$ 
as follows,
\begin{eqnarray} 
\bar\Gamma &=& \Gamma_{int}\Gamma_{ext}\\
&=&
\left(
\begin{array}{cc}
1 &  \\ 
& -1
\end{array}
\right)
\end{eqnarray}
where $\Gamma_{int}$ and $\Gamma_{ext}$ are given by 
\begin{eqnarray}
\Gamma_{int}&=&(-i)^{n}\Gamma_1\cdots\Gamma_{2n}\\
&=& 
\left(
\begin{array}{cc}
\gamma_{2n+1} &  \\ 
 & \gamma_{2n+1}
\end{array}
\right) ,\\
\Gamma_{ext}&=&(-i)\Gamma_{2n+1}\Gamma_{2n+2}\\
&=&
\left(
\begin{array}{cc}
\gamma_{2n+1} & \\ 
& -\gamma_{2n+1}
\end{array}
\right).
\end{eqnarray}

\section{General hermitian action on the spider's web lattice}
In this appendix 
we show the general form of two actions 
$S_{naive}$ and $S_W$ 
without any boundary conditions or constraints such as 
$\psi(\rho=1,\phi)=\psi(\rho=0)$ 
introduced in section 3.
The actions, which are constructed so that they 
have the hermiticity 
in both $\rho$ and $\phi$ directions
on the spider's web lattice, 
include a free parameter respectively
which should be fixed properly. 
We discuss another method of parameter fixing 
besides the constraint introduced in section 3. 

The naive action is given by 
\begin{equation}
  S_{naive}(b)=\sum_x \sum_\phi \sum_\rho \rho a_\rho 
\left[\bar\psi \Gamma_\mu \nabla_\mu \psi(b) + 
\bar\psi (\Phi_1 + i\bar\Gamma\Phi_2) \psi \right]
\end{equation}
where
\begin{equation}
\sum_\phi \sum_\rho \rho a_\rho \bar\psi 
\Gamma_\mu \nabla_\mu \psi (b) = 
\sum_\phi \sum_\rho \rho a_\rho \bar\psi 
(\Gamma_\mu \nabla_\mu)_{int}\psi 
+\sum_\phi \sum_\rho \rho a_\rho \bar\psi 
(\Gamma_\mu \nabla_\mu)_{ext}\psi(b) \quad ,
\end{equation}
\begin{eqnarray}
&&\sum_\rho \sum_\phi \rho a_\rho 
\bar\psi(\Gamma_\mu \nabla_\mu)_{ext}\psi(b) \nonumber\\
&=& 
b\sum_\phi \left[\bar\psi_{0,\phi} \Gamma_{2n+1}
\left(\cos\phi +i\Gamma_{ext}\sin\phi \right)
\frac{\psi_{1,\phi}}{2}-
\bar\psi_{1,\phi} \Gamma_{2n+1}
\left(\cos\phi +i\Gamma_{ext}\sin\phi \right)
\frac{\psi_{0,\phi}}{2}
\right] \nonumber \\
&&-\sum_\phi \bar\psi_{0,\phi} \Gamma_{2n+1}\frac{1}{2}
\left[\left(\sin(\phi+\frac{a_\phi}{2})\nabla_\phi + 
\sin(\phi-\frac{a_\phi}{2})\nabla_\phi^*\right) \right.
\nonumber \\
&&\left.-i\Gamma_{ext} 
\left(\cos(\phi+\frac{a_\phi}{2})\nabla_\phi + 
\cos(\phi-\frac{a_\phi}{2})\nabla_\phi^*\right)
 \right] \frac{\psi_{1,\phi}}{2} \nonumber\\
&&+\sum_{\rho=1}^{\infty}\sum_{\phi}\rho a_\rho 
\bar\psi_{\rho,\phi}\Gamma_{2n+1}
\left[\left(\cos\phi + i\Gamma_{ext}\sin\phi \right)\nabla_{\rho}^h 
\psi_{\rho,\phi} \right. \nonumber\\
&&-\frac{1}{2\rho a_\rho} 
\left\{  \left(\sin(\phi+\frac{a_\phi}{2})  
-i\Gamma_{ext}\cos(\phi+\frac{a_\phi}{2})\right)
\nabla_\phi \right. \nonumber \\ 
&&\left.+ \left(\sin(\phi-\frac{a_\phi}{2})  
-i\Gamma_{ext}\cos(\phi-\frac{a_\phi}{2})\right)\nabla_\phi^* 
\right\}
\left. \times 
\left(\frac{\psi_{\rho+1,\phi}+\psi_{\rho-1,\phi}}{2} \right)
\right] \label{axparameter_b}
\end{eqnarray}
and
\begin{eqnarray}
f(\rho)&=&m_0 \theta(\rho)\equiv 
\frac{\sinh(a_\rho\nu)}{a_\rho} \theta(\rho) \quad,\\
\theta(\rho)&\equiv& \left\{
\begin{array}{ll}
1 & \rho\geq 1 \\
0 & \rho =0 \quad .
\end{array}
\right.  
\end{eqnarray}
We note here that $S_{naive}(b)$ contains a free parameter $b$ 
in eq.(\ref{axparameter_b}) which 
should be fixed to an appropriate value in each model 
we construct.
This means that the action has some ambiguity at $\rho=0$ 
boundary. 
If we impose the constraint 
$\psi(\rho=1,\phi)=\psi(\rho=0)$ and 
redefine $\Phi$, we obtain the 
action $S_{naive}^c$ of eq.(\ref{S_naive^c}) 
in section 3. 

We now consider another example of parameter fixing.
Starting from the action $S_{naive}(b)$, 
we obtain the following zero mode solution for non-zero $b$, 
\begin{equation}
\varphi_-(2)= (b+1)\varphi_-(0), \quad \varphi_-(1)=0,
\end{equation}
and for $\rho \geq 2$,
\begin{equation}
\varphi_-(\rho+1)=-2f(\rho)a_\rho\varphi_-(\rho)
+\varphi_-(\rho-1). 
\end{equation}
This solution exists for $k=0$. 
We see that there exists a zero mode solution 
for $b$ which satisfies $b\neq0,-1$ and 
which is not too large.
In any case we need the extended Wilson term since 
there appear doubling species.

Next we shall see the general form of extended Wilson term.
The extended Wilson term is given by, 
\begin{eqnarray}
S_W(c) &=& S_W^{in} + S_W^{out}(c) ,\\
S_W^{in}&=&\sum_x \sum_{\rho=0}^{\infty}\sum_{\phi}
\frac{w a}{m_0}
\sum_{i=1}^{2n} \bar\psi_{\rho,\phi} \rho a_\rho
(\Phi_1 +i\bar\Gamma\Phi_2)
\Delta_i \psi_{\rho,\phi} ,\\
S_W^{out}(c)&=&\sum_x \sum_{\rho=1}^{\infty}
\sum_{\phi} \frac{w_{\rho}a_{\rho}}{m_0}
\bar\psi_{\rho,\phi} \rho a_\rho
\left[(\Phi_1 +i\bar\Gamma\Phi_2)
\Delta_\rho \psi_{\rho,\phi}
+\Delta_\rho \left((\Phi_1 +i\bar\Gamma\Phi_2)\psi_{\rho,\phi}\right)
\right]\times\frac{1}{2} \nonumber \\
&+&\sum_x \sum_{\rho=1}^{\infty}\sum_{\phi}
\frac{w_{\rho}a_{\rho}}{m_0}
\bar\psi_{\rho,\phi} 
\left[(\Phi_1 +i\bar\Gamma\Phi_2)
\nabla_\rho^h \psi_{\rho,\phi}
+\nabla_\rho^h \left((\Phi_1 +i\bar\Gamma\Phi_2)\psi_{\rho,\phi}\right)
\right]\times\frac{1}{2} \nonumber \\
&+&\sum_x \sum_{\phi} \frac{w_{\rho}}{m_0}
\left[c\bar\psi_0 \left(\frac{f_0+f_1}{2}\right)
e^{i\bar\Gamma \phi} \psi_1 
+(c-\frac{1}{2})\bar\psi_1 \left(\frac{f_0+f_1}{2}\right)
e^{i\bar\Gamma \phi} \psi_0 
\right]  \nonumber \\
&+&\sum_x \sum_{\rho=0}^{\infty}\sum_{\phi}
\frac{w_\phi a_\phi^2 a_{\rho}}{m_0}
\bar\psi_{\rho,\phi}\frac{1}{\rho a_\rho}
\left[(\Phi_1 +i\bar\Gamma\Phi_2)
\Delta_\phi \psi_{\rho,\phi}
+\Delta_\phi \left((\Phi_1 +i\bar\Gamma\Phi_2)\psi_{\rho,\phi}\right)
\right]\times\frac{1}{2} \label{axparameter_c} \nonumber\\
\end{eqnarray}
where $c$ is a free parameter like $b$ 
in the naive action $S_{naive}(b)$. 
$S_W(c)$ has some ambiguity 
at $\rho=0$ boundary, too. 
We have to adjust this parameter so that 
the model has a good property.
If we impose the constraint 
$\psi(\rho=1,\phi)=\psi(\rho=0)$ and 
redefine $\Phi$, 
we obtain the action $S_W^c$ of eq.(\ref{S_W^c}) 
in section 3. 

We now consider another example of parameter fixing besides 
the constraint $\psi(\rho=1,\phi)=\psi(\rho=0)$. 
We start from the action where the parameters 
are fixed as follows, 
\begin{equation}
S=S_{naive}(b=-w_\rho )+S_W(c=1). \label{axpara_bc}
\end{equation}
Solving the zero mode equations,
we obtain the zero mode solution
\begin{equation}
\varphi_-(2)=
\frac{4}{5}\left[1-a_\rho m_0 
\left(1-\frac{1}{4 a_\rho m_0}\right) +F
\right]\varphi_-(1) ,  \label{axphi_2}
\end{equation}
and for $\rho \geq 2$,
\begin{equation}
\varphi_-(\rho+1)=
\frac{\left[1-a_\rho m_0\left(1-\frac{1}{4 \rho^2 a_\rho m_0}\right) +F \right]
\varphi_-(\rho)+\frac{1}{4\rho}\varphi_-(\rho-1)}
{1+\frac{1}{4\rho}}, \label{axphi_rho}
\end{equation}
where we have chosen $w_\rho$, $w_\phi$ and $w$ such that 
$w_\rho=w_\phi a_\phi^2 = \frac{1}{2}$ and $\frac{w a_\rho}{a}= \frac{1}{2}$.
This solution exists for $k=0$. 
From eqs.(\ref{axphi_2})
(\ref{axphi_rho}), 
we find that 
the input of $\varphi_-(1)$ 
allows us to obtain a unique $\varphi_-(\rho)$ 
from $\rho=2$ to $\rho=\infty$ 
inductively. 
$\varphi_-(0)$ does not appear in the zero mode 
equations, which means that $\varphi_-(0)$ 
is not a dynamical field.\footnote{In this case 
the string has 
a finite thickness whose radius is about $a_\rho$ and that 
the area within one site from $\rho=0$ is excluded 
from the spider's web lattice.}
We see that there exists a 
normalizable\footnote{The normalizability 
is satisfied since the discussion 
below eqs.(\ref{p_rho})(\ref{p_rhoz}) in section 3 
is repeated.} zero mode
solution for $0<am_0<2$ also in this 
parameter fixing.

However, this parameter fixing has some problem. 
When we try to obtain $4$ dimensional low energy 
effective theory explicitly,  
the procedure discussed as eq.(\ref{subtraction}) in 
section 3 may become a little more complicated 
because we need to consider some average of $\phi$ at the 
site $(\rho=1,\phi)$.
On the other hand, we do not have to care about this 
with the constraint $\psi(\rho=1,\phi)=\psi(\rho=0)$. 

%
%%%%%%%%%%%%%%%%%%%%%%%%%%%%%%%%%%%%%%%%%%%%%%%%%%%%%%%%%%%%%%%%

%%%%%%%%%%%%%%%%%%%%%%%%%%%%%%%%%%%%%%%%%%%%%%%%%%%%%%%%%%%%%%%%

%

\end{document}